\documentclass[10pt]{article}
\usepackage{amsmath}
\usepackage{amsfonts}

\begin{document}
\title{Requirement for quantum computation} 
\author{Stephen D.\ Bartlett and Barry C.\ Sanders \\
  Australian Centre for Quantum Computer Technology, \\
  Department of Physics, Macquarie University, \\
  Sydney, New South Wales 2109, Australia} 
\date{16 April 2003}

\maketitle 
\begin{abstract}
  We identify ``proper quantum computation'' with computational
  processes that cannot be efficiently simulated on a classical
  computer.  For optical quantum computation, we establish no-go
  theorems for classes of quantum optical experiments that cannot
  yield proper quantum computation, and we identify requirements for
  optical proper quantum computation that correspond to violations of
  assumptions underpinning the no-go theorems.
\end{abstract}

\section{Introduction}

Quantum computation~\cite{Fey82,Nie00} offers the possibility of
(i)~efficiently simulating quantum dynamics for which classical
simulations are hard and (ii)~solving computational problems for which
no efficient algorithm is known for classical computation.  In the
latter case, Shor's algorithm for efficient factorization~\cite{Sho95}
is a famous example of the potential of quantum computation, which has
motivated a global effort to develop quantum computers.  As quantum
mechanics underpins all physical theories, it is important to identify
the requirements for ``proper quantum computation,'' where we use this
term to refer to quantum computations that outperform those allowed by
a strict application of the laws of classical physics.

A key issue in the design of these quantum computers is identifying
requirements such as gates with entanglement capability and
appropriate sources and detectors that can perform tasks not possible
with a classical machine.  An important question is: How do we know
when a quantum process may yield a computational advantage?

One approach to identifying quantum processes that may potentially
lead to quantum computation is to identify what \emph{is not} proper
quantum computation.  To address this question in a quantum optics
setting, we have identified a large ``toolkit'' consisting of sources,
processors and detectors such that any network consisting of these
devices can be efficiently simulated on a classical computer.  With
this toolkit, we construct no-go theorems that elucidate requirements
for quantum computation.  On one hand, a surprisingly large class of
quantum networks are shown not to allow for quantum computation in the
sense of solving problems that are intractable on a classical machine.
On the other hand, the critical resources for performing quantum
computation are exposed as violations of assumptions in the theorems.
Identifying resources through violations of assumptions is
particularly useful in understanding schemes that eliminate the
apparent requirement of an optical nonlinearity by utilizing certain
measurements.  Our no-go theorems also suggest how stringent criteria
might be relaxed and still be able to deliver powerful quantum
computation.
  
\section{Computing: problems and algorithms}

Quantum computation is often regarded as a powerful, ``non-classical''
computation, i.e., a computation that cannot be performed ``easily''
on a conventional computer.  However, this description requires us to
be specific about what we mean by a computation, by ``non-classical'',
and by computationally ``easy''.  Because we expect a quantum computer
to be able to do anything that a classical computer can do (and
possibly much more), we need to rule out classes of computations on a
quantum computer that are not \emph{proper}, in the sense that they
could equally well be performed on a classical computer.  In this
section, we define relevant computing concepts and terms with the aim
of identifying what constitutes proper quantum computation.

A \emph{computational problem} is a mathematical function that maps an
\emph{instance} (i.e., an input) to a \emph{solution}.  The sets of
instances and solutions are classical sets of distinguishable
elements, e.g., integer numbers.  For example, the problem, ``Is an
integer $p$ a prime number?'' is a computational problem, one for
which the set of instances is the set of integer numbers and the set
of solutions consists of the logical output ``yes (it is prime)'' and
``no (it is not)''.

For a specific computational problem, an \emph{algorithm} is a
detailed, step-by-step method or recipe for finding the solution
corresponding to a given instance~\cite{Pap94}, using an agreed set of
operations.  These operations may include boolean operations or
unitary quantum transformations or something else.  An algorithm is
thus a prescription, using specified operations, for solving the
problem in general.  Finding the solution for a given instance simply
becomes a mechanistic process: follow the steps of the algorithm to
obtain the solution.

Once an algorithm for a problem is devised, the process of calculating
solutions for various instances can be automated.  A \emph{computer}
is a physical, mechanistic device used to implement the steps of an
algorithm, thus calculating a solution for a given instance of the
problem.  A computer is mechanistic and employs specific operations,
and it is important to identify the physical laws governing the
computer's operations; different physical laws underpinning the
computation could affect the computer's capabilities.  We define a
\emph{classical computer} as a computer whose architecture obeys the
laws of classical physics.  This point is rather subtle: the
architecture of existing computers is classical because it \emph{can},
in principle, be implemented in systems that obey the laws of
classical physics; in reality devices such as transistors, which are
described as quantum devices, does not render such architectures
quantal because quantum devices are a technical convenience, not an in
principle requirement.

The definition of a classical computer is now clear and can be
extended to define the quantum computer.  A \emph{quantum computer} is
one whose processes obey the laws of quantum physics.  Because quantum
physics is fundamental (and classical physics emerges from quantum
laws in particular circumstances), our definition leads to a classical
computer being a special case of a quantum computer, one for which the
governing natural laws are restricted to the limit where classical
physics applies.

For a specified computation problem, the chosen algorithm determines
if a quantum computer is required or if a classical computer will
suffice.  If the algorithm employs classical operations, a classical
computer suffices; if quantum operations are employed, then a quantum
computer is needed (although of course a classical computer may be
able to simulate the quantum algorithm).  Thus, we define a
\emph{quantum algorithm} as one that requires quantum operations to
implement.  In contrast, a classical computer program is a
\emph{classical algorithm} that requires only classical operations for
its execution.

With these definitions, we can address the question, ``What is a
quantum computation?''  It seems natural to define quantum computation
as a physical process, obeying the laws of quantum physics, that
provides a solution for a given instance of a computational problem.
This definition is unsatisfactory for our purposes: any computation on
a conventional computer (which fundamentally obeys the laws of quantum
physics) would thus be considered a quantum computation.

A better definition might be obtained by demanding that demonstratable
quantum dynamics (such as quantum tunnelling, quantum superpositions
or entanglement) occur during the computation. This approach is
appealing because one could experimentally test for quantum effects
and claim that a quantum computation is occurring. However, the
problem with this extended definition is that the presence of quantum
effects may not yield a quantum computation that performs better than
a classical computation. Quantum effects could enhance the performance
of computation, but incorporating entanglement does not guarantee this
performance enhancement. The definition needs to consider the
performance of the computation.

Proper quantum computation may be described as computation that
outperforms comparable (or possibly any) classical computation.  In
order to quantify this concept of outperformance, we appeal to the
subject of computational complexity.

\subsection{Computational complexity}

Quantum computation, per se, had its origins in Feynman's
musings~\cite{Fey82}, motivated by using quantum processes to
efficiently simulate quantum dynamics for cases where classical
computers appear to be grossly inadequate.  The issues of
\emph{efficiency} and \emph{efficient simulation} are an active area
of research in computer science known as computational
complexity~\cite{Pap94}, and will be useful in defining a concept of
proper quantum computation.

The motivation of computational complexity is to quantify the
difficulty of an algorithm for a computational problem.  Note that
``difficulty'' does not quantify how challenging it is for a
researcher to devise an algorithm for a problem, but instead
quantifies the physical resources that are required by a computer to
implement an algorithm for a given instance size.  More precisely, it
is a comparison of the amount of physical resources $R$ (e.g., the
number of computational steps, or the amount of physical memory)
required in order to obtain a solution for an instance with size $S$
(e.g., the number of bits needed to write the instance into memory).

The computational complexity class $\mathbf{P}$ is, loosely, the set
of algorithms for which $R$ is bounded by a polynomial function of
$S$.  For such an algorithm, it is often said that the amount of
physical resources required to solve the problem ``scales
polynomially'' in the size of the instance.  Classical algorithms that
are in $\mathbf{P}$ are generally considered to be easy, or
\emph{efficient}, whereas an algorithm that is not in $\mathbf{P}$ is
considered hard and inefficient.  Note that this classification is a
loose guide (an algorithm where $R$ behaves as $S^{100}$ is in
$\mathbf{P}$ but may be practically intractable), but is one that in
practice serves as an excellent classification of easy and hard
algorithms.

It could be argued that the goal of quantum computer science is to
devise quantum algorithms for which the number of quantum operations
required scales polynomially in the size of the instance, but where no
classical algorithm in $\mathbf{P}$ is known or even possible.  In
such cases, allowing for quantum operations clearly results in an
advantage from a computational perspective.  Shor's quantum algorithm
for efficient factorization is one such example; finding more examples
may lead to an understanding of how quantum computation may be more
powerful than classical.

\subsection{Classical simulation of quantum processes}

As discussed in the previous section, allowing quantum operations in a
computation does not guarantee an advantage from a computational
complexity perspective.  In order to identify requirements for proper
quantum computation as distinct from classical computing, we
require that proper quantum computation offers an advantage
over classical computing.  Consider a quantum algorithm, or even more
generally, a quantum process.  We say that this process can be
\emph{efficiently simulated} on a classical computer if the classical
computing resources required to simulate it scale polynomially in the
size of the quantum process; i.e., if the classical algorithm
simulating the quantum one is in $\mathbf{P}$.  The concept of
efficient classical simulation leads us to the following definition.

\textbf{Definition (proper quantum computation):}  A quantum
computation that cannot be efficiently simulated on a classical
computer.

We immediately see that this definition is unsatisfactory in some
respects: some quantum computations offer a polynomial speedup over
classical algorithms (such as Grover's search algorithm~\cite{Gro97},
which offers a quadratic speedup over classical searches).  These
computations can be efficiently classically simulated and are thus not
proper quantum computations by our definition; nevertheless they still
qualify as quantum algorithms.  However, quantum computations that are
proper (and thus not efficiently classically simulatable) are of
paramount importance, and thus we focus our attention on identifying
requirements for them.

Proving that a quantum algorithm is proper and will outperform
\emph{any} classical algorithm is difficult.  One fruitful
direction towards this goal is to identify large classes of quantum
processes that do \emph{not} offer an advantage over a classical
system: this exclusive direction serves to focus the search for proper
quantum algorithms.  Such efficient simulation does not imply that
quantum effects are not present in the quantum system, but simply
notes that the quantum system does not provide a computational
advantage.  Appealing to Feynman's original concept, the interesting
applications for a quantum computer are for quantum processes that
\emph{cannot} be efficiently simulated on a classical computer.

\section{Optical realizations of quantum computation}

We now turn our attention to performing quantum computation with
quantum optics.  Optical realizations of quantum information
processing benefit from advanced techniques in quantum optics for
state preparation, unitary evolution with low decoherence and
high-efficiency measurement.  Both qubit~\cite{Chu95,KLM01,Got01b} and
continuous-variable~\cite{Llo99,Bar03} schemes allow optical quantum
information processing; experiments demonstrating optical quantum
teleportation~\cite{Bou97,Fur98,Pan03} and proposals for schemes such
as optical quantum secret sharing~\cite{Tyc02,Lan03} are testimony to
advances in quantum optical quantum information tasks and processes.

It is important to determine the useful and necessary optical
processes to perform proper quantum computation, and we specifically
identify classes of processes that can be efficiently simulated on a
classical computer.  The Gottesman-Knill (GK)
theorem~\cite{Got99,Nie00} for qubits shows that it is sometimes
possible to efficiently simulate a restricted set of quantum
operations on a classical computer via a clever representation.  In
the following, we construct a large toolkit of quantum optics sources,
processors and detectors that can be efficiently simulated on a
classical computer.  This construction allows us to identify key
resources outside of this toolkit that may allow for proper quantum
computation.

\subsection{Simulating optical quantum processes}

A full quantal treatment of a mode of the electromagnetic field
requires the infinite-dimensional Hilbert space of a harmonic
oscillator.  Clearly, attempting to represent the state of many
coupled optical modes on a classical computer is a daunting task due
to the shear size of the quantum Hilbert space.  Only more compact
representations of a restricted set of such quantum states and
corresponding transformations could ever be made tractable on a
classical computer.  We demonstrate in the following that such compact
representations exist for a wide range of optical quantum networks.

Linear optics, consisting of beam splitters, phase-shifters and other
linear couplers together with semiclassical sources and coherence
measurements can be described purely through a semiclassical
description, and thus cannot yield proper quantum computation because
the linear semiclassical evolution can be efficiently simulated.
Also, single-photon schemes that employ only linear
optics~\cite{Cer98} are not scalable, in that they require resources
that grow exponentially in the number of qubits~\cite{Blu02}.
Squeezing processes are realised by a $\chi^{(2)}$ nonlinearity; if
the pump is treated classically using a mean-field approximation, the
resulting operation can be viewed as linear on a single mode (one-mode
squeezing) or two modes (yielding an entangling transformation).  The
addition of squeezing and entangling transformations to this scheme is
also insufficient, as proven by the continuous-variable classical
simulatability theorem of Bartlett \emph{et al}.~\cite{Bar02}:

\medskip
\noindent \textbf{Theorem 1 (Efficient Classical Simulation of
  Continuous Variable Quantum Information):} \textit{Any continuous
  variable quantum information process that initiates with Gaussian
  product states (products of squeezed displaced vacuum states) and
  performs only (i) linear phase-space displacements, (ii) squeezing
  transformations on a single oscillator mode, (iii) two-mode
  squeezing transformations, (iv) measurements of quadrature phase
  (i.e., homodyne detection) with finite losses, and (v) any such
  operations conditioned on classical numbers or homodyne detection
  (classical feed-forward), can be \emph{efficiently} simulated using
  a classical computer.}  \medskip

An outline of the proof of this theorem is as follows.  Gaussian
states of an $N$-mode optical system are completely characterised by
the vector consisting of the mean values of the canonical variables
and by the covariance matrix.  This representation of the states can
be stored efficiently on a classical computer.  Linear optics and one-
and two-mode squeezing transformations possess a straightforward group
action on this representation: these operations displace the means and
transform the covariance matrix but maintain the Gaussian property of
the multi-mode state.  Because of the ease of this representation,
these calculations can be simulated efficiently on a classical
computer.  

We note that the inclusion of squeezing into this list allows for
non-Poissonian photon statistics.  For example, the output of
parametric downconversion is described by a state with only even
photon number contributions.  Such squeezed states, however, are still
Gaussian and fall within the constraints of our theorem.  Thus,
although techniques of linear optics and squeezing with semiclassical
sources and homodyne detection are highly advanced and can demonstrate
non-classical properties such as quantum
teleportation~\cite{Bou97,Fur98,Pan03} and quantum secret
sharing~\cite{Tyc02,Lan03}, they are insufficient to perform proper
quantum computation.

Recently, non-unitary processes such as measurement have been
identified as a means to extend the power of optical quantum
information processing~\cite{KLM01,Got01b}.  The essence of such
schemes is that two optical systems (e.g., modes) are entangled,
followed by a measurement on one system.  The other system
``collapses'' into a state that depends on the measurement outcome;
for certain outcomes this collapse can be seen as equivalent to a
unitary transformation.  Proposals by Knill, Laflamme and
Milburn~\cite{KLM01} (KLM) and Gottesman, Kitaev and
Preskill~\cite{Got01b} (GKP) employ photon counting to induce unitary
transformations in optical systems non-deterministically (i.e., they
occur when certain measurement outcomes are observed), which leads to
potential experimental schemes for optical quantum computation.

The above classical simulatability theorem can be extended to include
non-unitary processes such as measurement.  The classical
simulatability theorem of Bartlett and Sanders~\cite{Bar02c} employs
the powerful formalism of Gaussian completely positive (CP)
maps~\cite{Lin00} to describe efficiently simulatable operations
(including some non-unitary processes such as measurement) on Gaussian
states.

\medskip
\noindent \textbf{Theorem 2 (Efficient Classical Simulation of
  Optical Processes):} \textit{Any quantum information process that
  initiates in a Gaussian state and that performs only Gaussian CP
  maps can be \emph{efficiently} simulated using a classical
  computer.}  These maps include (i) the unitary transformations
corresponding to linear optics and squeezing, (ii) linear
amplification (including phase-insensitive and phase-sensitive
amplification and optimal cloning), linear loss mechanisms or additive
noise, (iii) measurements that are Gaussian CP maps including, but not
limited to, projective measurements in the position/momentum
eigenstate basis or coherent/squeezed state basis, with finite losses,
and (iv) any of the above Gaussian CP maps conditioned on classical
numbers or the outcomes of prior Gaussian CP measurements (classical
feedforward).  \medskip

Again, the proof of this theorem lies in the simple representation for
Gaussian states given by the means and covariance matrix.  The
non-unitary operations covered by this theorem form a semigroup
(similar to a group but without the guarantee that every element is
invertible) that again preserve the Gaussian nature of the states.
This theorem for efficient classical simulation provides a powerful
tool in assessing whether a given optical process can enhance linear
optics to allow for proper quantum computation.  Algorithms or
circuits employing Gaussian-preserving maps can be efficiently
simulated on a classical computer, and thus cannot lead to proper
quantum computation.

The results of this section reveal the limitation of using Gaussian
states for quantum information processing: the existence of a compact
representation for these states (and Gaussian-preserving
transformations and measurements on them) leads to no-go theorems for
proper quantum computation.  Clearly, ``going beyond'' Gaussian states
in optical quantum computation is necessary (although possibly not
sufficient).

\subsection{Requirement of optical nonlinearity}

In particular, higher-order optical nonlinear processes (such as a
Kerr nonlinearity~\cite{Wal94}), which can yield non-Gaussian states,
have been identified as a necessary requirement~\cite{Llo99,Bar02} for
proper quantum computation with optics.  Unfortunately, Kerr
nonlinearities suffer either from weak strengths or high losses, and
the lack of appropriate nonlinear materials greatly restricts the type
of processes that can be performed in practice.  Optical quantum
computation schemes such as~\cite{KLM01,Got01b} use measurements to
induce a nonlinear transformation; however, as shown in the previous
theorems, not all forms of measurement can yield proper quantum
computation.  In this section, we discuss measurements that may be
used to induce an optical nonlinearty and those which cannot by
employing the results of our no-go theorems.

First, our theorem provides a strong no-go result for the use of
homodyne measurement:

\medskip
\noindent \textbf{Corollary 1:}  \emph{Linear optics or squeezing
transformations conditioned on the measurement outcome of homodyne
detection with finite losses using Gaussian states cannot induce a
nonlinearity.}  \medskip

Thus, initiating with Gaussian states, it is not possible to use
homodyne measurements and feedforward of measurement results to induce
a (possibly nondeterministic) optical nonlinearity in the way that
photon counting allows in the KLM scheme.  In terms of optical
implementations of quantum computing, this theorem reveals why all
previous schemes either propose some form of optical
nonlinearity~\cite{Chu95,Llo99}, use other forms of measurement such
as photon counting~\cite{KLM01,Got01b} or are not efficiently
scalable~\cite{Cer98}.

This theorem also places severe constraints on the use of
photodetection to perform nonlinear transformations in realizations of
optical quantum computing.  For a threshold
photodetector~\cite{KLM01,Kok01,Bar02b} with perfect efficiency, the
POVM is given by two elements, corresponding to ``absorption'' and
``no-absorption'' of light.  Photon counters are effectively
constructed as arrays of such detectors~\cite{Bar02b}.  The vacuum
projection describes the non-absorption measurement, and the
corresponding map describing this measurement result is Gaussian CP.
However, the absorption outcome is not.

\medskip
\noindent \textbf{Corollary 2:}  \emph{Gaussian-preserving maps
  conditioned on the no-absorption outcome of a photodetection
  measurement can be efficiently simulated on a classical computer;
  transformations conditioned on the absorption outcome
  cannot be efficiently simulated in this manner.}  \medskip

Note that the same result holds for finite-efficiency photodetectors:
such detectors can be modelled as unit efficiency photodetectors with
a linear loss mechanism describable using Gaussian CP maps.  Thus, the
absorption outcome of photodetection and the feedforward of this
measurement result is a key resource for optical quantum information
processing.  This corollary also proves that any nonlinear gate
employing linear optics and photon counting \emph{must} be
nondeterministic; a photon counting measurement of a Gaussian state
could possibly result in an outcome of zero photons, and such a result
corresponds to an efficient, classically simulatable process.  (Note
that nonlinear optics, in contrast, allows deterministic processing.)

\begin{table}[t]
\caption{Efficient classical simulatability for schemes employing
  various initial states, unitary gates, and measurements.\label{tab:Sim}}
\begin{center}
\footnotesize
\begin{tabular}{|l|l|l|l|}
\hline
{\bf Initial States} &\raisebox{0pt}[13pt][7pt]{\bf Unitary Gates} &
\raisebox{0pt}[13pt][7pt]{\bf Measurements} & 
\begin{minipage}{0.7in}{\bf Efficiently}\\{\bf simulatable}
\end{minipage}\\
\hline
{Vacua} &\raisebox{0pt}[13pt][7pt]{Linear optics, squeezing} &
\begin{minipage}{1in}
{Gaussian CP}\\ {(i.e., homodyne)}
\end{minipage} &{\checkmark~\cite{Bar02,Bar02c}}\\
\hline
{Vacua} &\begin{minipage}{1.25in}{Linear optics, squeezing,}\\ {Kerr
    nonlinearity} \end{minipage} &
\raisebox{0pt}[13pt][7pt]{Homodyne} &
\begin{minipage}{0.7in}
  {$\times$~\cite{Llo99}}
\end{minipage}\\
\hline
{Single photons} &\raisebox{0pt}[13pt][7pt]{Linear optics only} &
\raisebox{0pt}[13pt][7pt]{Photon counting} &
\raisebox{0pt}[13pt][7pt]{$\times$~\cite{KLM01}}\\
\hline
{Vacua} &\raisebox{0pt}[13pt][7pt]{Linear optics, squeezing} &
\begin{minipage}{1in}{Photon counting}\\{\& homodyne} 
\end{minipage}&
\raisebox{0pt}[13pt][7pt]{$\times$~\cite{Got01b}}\\
\hline
{Single photons} &\raisebox{0pt}[13pt][7pt]{Linear optics, squeezing} &
\raisebox{0pt}[13pt][7pt]{Homodyne} &{?}\\
\hline
\end{tabular}
\end{center}
\end{table}

Our classical simulatability theorem may be useful in assessing the
minimum requirements for proper quantum computation with optics.
Table 1 presents various classes of initial states, unitary gates, and
measurements (that can be used for classical feedforward) and their
classical simulatability according to our theorem.  Employing only
Gaussian states and Gaussian CP maps results in an efficiently
simulatable circuit; one can now consider supplementing this set with
various ``resources'' that may allow for proper quantum computation.
As shown by Lloyd and Braunstein~\cite{Llo99}, the addition of a Kerr
nonlinearity or any higher-order transformation on a single mode
results in the ability to efficiently simulate the evolution of any
polynomial Hamiltonian.  The schemes of KLM and GKP reveal that photon
counting is also a resource that allows for universal quantum
computation.  The KLM scheme also requires single photon Fock states
``on demand'' as ancilla inputs to their nondeterministic nonlinear
gates; such states lie outside the domain of our theorem (they are not
Gaussian) and may serve as a resource for performing nonlinear
operations.

It is interesting to consider, then, if single photons on demand are
by themselves sufficient to bestow Gaussian CP maps with the power to
perform nonlinear operations and thus possibly proper quantum
computation.  Considering the recent progress in creating single
photon turnstile devices~\cite{San01} (with low probablility of
producing zero or two photons by accident), a scheme that requires
single photons but otherwise employs only linear optics, squeezing,
and high-efficiency homodyne detection would obviate the need for
ultra-high efficiency photon counters~\cite{Bar02d}.

\section{Discussion}

Quantum optics is challenged by the advent of quantum computation, not
only by the technical hurdles that must be overcome to achieve
scalable quantum computers, but also by the fundamental question of
creating and verifying proper quantum computation in the laboratory.
Here we have elucidated the nature of quantum computation, pointed out
that a large quantum optics toolkit is insufficient to realise proper
quantum computation and illustrated how violating assumptions in our
no-go theorems may correspond to requirements for quantum computation.

We have established that optical transformations that map Gaussian
wavefunctions into Gaussian wavefunctions, whether unitary or not, are
insufficient to perform proper quantum computation.  If the Gaussian
nature of wavefunctions is preserved by the operations, then we can
exploit the mean-and-covariance representation of states to implement
an efficient classical simulation of the quantum system; hence this
quantum computation does not sufficiently outperform a classical
computation to qualify as proper quantum computation according to our
criteria.

We highlight the importance of non-Gaussian transformations, whether
they correspond to nonlinear unitary evolutions or to the nonlinear
CP-map of conditioning a unitary transformation on detecting photons
(the absorption outcome), and also the initiation with non-Gaussian
states, to violate our theorems. If the initial states are
non-Gaussian, or if the transformations do not preserve the Gaussian
representation of wavefunctions, then the ``clever'' representation of
Gaussian states in terms of means and variances is no longer adequate
to provide algorithms for efficient classical simulation. Although we
cannot guarantee that violating our theorem is enough for proper
quantum computation, we can -- and do -- rule out classes of
experiments (as determined by the ``toolkit'' in the laboratory) as
being sufficient for proper quantum computation.

In designing experiments for quantum computation, one must always
consider the most efficient means to simulate the processes and
outcomes on a classical computer. If a clever means exists to simulate
the experimental quantum computation on a classical computer such that
this simulation is in \textbf{P}, then the quantum computation is not
\emph{proper}: our requirement is that the toolkit must be sufficient
to produce quantum computations that defy efficient classical
simulation.

\subsection*{Acknowledgments}

This project has been supported by the Australian Research Council and
Macquarie University.

\end{document}